\documentstyle[prl,aps,multicol,epsfig]{revtex}
\begin{document}
\draft
%%%%%%%%%%%%%%%%%%%%%%%%%%%%%%%%%%%%%%%%%%%%%%%%%
\newcommand{\be}{\begin{equation}}
\newcommand{\ee}{\end{equation}}
\newcommand{\ba}{\begin{eqnarray}}
\newcommand{\ea}{\end{eqnarray}}
\newcommand{\vk}{{\bf k}}
\newcommand{\vq}{{\bf q}}
%%%%%%%%%%%%%%%%%%%%%%%%%%%%%%%%%%%%%%%%%%%%%%%%%%%%%%%%%%%%%%%%%%%%%5

\title{Resonant Raman Scattering of Interacting Two-Channel Quantum Wires}
\author{Hyun C. Lee }
\address{BK21 Physics Research Division and Institute of Basic Science, 
Sung Kyun Kwan University, Suwon, 440-746
Korea}
\maketitle
\draft
\begin{abstract}
Resonant Raman scattering of degenerate interacting two-channel quantum wire
is studied.
{\it All} collective excitations of two-channel quantum wire are shown to
give rise to peaks in the polarized  Raman spectra near resonance. 
If there exist certain
 symmetries among interactions, a resonant peak can also appear in the
{\it depolarized} Raman spectra, in constrast to
the single-channel case studied by Sassetti and Kramer.  
We also calculate the explicit form
of  the scattering cross-section away from the peaks.  The above features
may be experimentally verified in armchair carbon nanotube systems.
\end{abstract}
\thispagestyle{empty}
\pacs{{\rm PACS numbers}: 71.27.+a, 71.10.pm, 78.30.-j}
\begin{multicols}{2}
%%%%%%%%%%%%%%%%%%%%%%%%%%%%%%%%%%%%%%%%%%%%%%%%%%%%%%%%%%%
\section{Introduction}
Raman spectroscopy is a very powerful tool in investigating 
the physical properties
of one-dimensional (1D) electron system.  
Especially, it can directly measure the dispersion of
collective excitations\cite{kramer,ramantheo}.
Away from resonance, depending on the relative polarizations of the incident 
and the scattered light, the charge density excitation (CDE, parallel
polarization, polarized spectra) 
and the spin density excitation (SDE, perpendicular polarization,
depolarized spectra)
can be identified in the Raman spectra \cite{kramer,ramantheo,ramanexp}.
Near resonance, {\it additional} structures have been detected 
in {\it both} polarized {\it and} depolarized spectra \cite{ramanexp},
and they were termed   "single particle excitation" (SPE).
Sassetti and Kramer pointed out that the SPE in
{\it polarized spectra} could be understood in terms of the spin density 
collective excitation in {\it depolarized spectra} 
within the theory of 1D interacting single-channel
electron gas in Luttinger liquid approximation.
They also predicted  a broad incoherent structure in {\it depolarized}
spectra near resonance, which appears due to the simultaneous propagation
of CDE and SDE (in other words, spin-charge separation).
Thus, the resonant Raman spectra can provide many experimental verifications
of anomalous non-Fermi liquid behaviours of 1D interacting electrons
\cite{Voit}.
%%%%%%%%%%%%%%%%%%%%%%%%%%%%%%%%%%%%%%%%%%%%%%%%%%%%%%%%%%%%%%

Two-channel quantum wires attracted much attention recently 
motivated by the carbon nanotubes  and the spin-ladder system \cite{nano,ladder}.
It is natural to explore the implications of Raman scattering on two-channel
quantum wire in the perspective of the above discussions.
We have extended the analysis by  Sassetti and Kramer to the 
{\it degenerate}(See Eq.(\ref{h0})) {\it two}-channel 
quantum wires, and have found some new features 
which are qualitatively different  from the single-channel ones.
Our findings are:
(a) {\it all} collective excitations of the system give rise to peaks in 
the resonant {\it polarized} Raman spectra. 
(b) If some of the collective excitations
have the identical velocitites owing to symmetries among interactions,
a {\it peak  appears} in the resonant {\it depolarized } Raman spectra, 
which is contrast to the prediction in the single-channel quantum wire.
Our predictions may be checked experimentally in the carbon nanotubes, 
as  will be shown below.
The above results should be compared with those by E. Mariani {\it et al.}.
\cite{mariani}
They have studied the resonant Raman spectra of {\it non-degenerate} 
two-channel quantum wires (i.e. a finite energy gap between two sub-bands).
Their results show that only collective inter-band spin excitations with positive
group velocity, apart from the charge excitation, can appear in resonant 
{\it polarized} spectra and that {\it only} broad features exist in resonant 
{\it depolarized} spectra. Because our system is degenerate there is no distinction
between intra- and inter-band modes apart from the renormalization due to
interactions.

In Section II, we describe a model of two-channel quantum wires and 
review some results in the standard Raman spectroscopy.
In Section III, we calculate the resonant Raman cross-section
and discuss the implications of the results.

%%%%%%%%%%%%%%%%%%%%%%%%%%%%%%%%%%%%%%%%%%%%%%%%%%%%%%%%%%%%%%%%%%%
\section{Model and Formulation}
As a representative model of the two-channel quantum wires, 
we study the "forward scattering charge model"
(FSCM) of $(N,N)$ armchair carbon nanotubes 
proposed by Kane, Balents, and Fisher\cite{kbf} and its slightly generalized 
model. FSCM neglects the backward and umklapp scatterings, and as a result
the Luttinger liquid description of FSCM 
holds above a certain gap scale \cite{kbf}.
Such a condition is implicitly assumed to be met.
We regard FSCM as a model of typical two-channel quantum wires
in Luttinger liquid universality class rather than 
as a microscopic model of carbon nanotubes.

With the above reservations, we describe the model.
A discrete quantized  transverse momenta leads to {\it two}
1D metallic bands with the {\it same} Fermi velocity.
Following the notations of Kane, Balents, 
the two metallic bands can be described by the Hamiltonian
\be
\label{h0}
H_0=\sum_{i,\alpha}\,\int dx v_F \Big[ \psi^\dag_{R i \alpha} i \partial_x 
\psi_{R i \alpha}-  \psi^\dag_{L i \alpha} i \partial_x 
\psi_{L i \alpha}\Big],
\ee
where $i=1,2$ labels the two bands, and $\alpha=\uparrow,\downarrow$ the 
electron spin.
The Hamiltonian (\ref{h0}) can be bosonized in a standard way\cite{Voit},
\be
H_0(\theta,\phi)= \sum_{i \alpha}\,
\int dx \frac{v_F}{2\pi}\,\Big[ (\partial_x \theta_{i \alpha})^2+
(\partial_x \phi_{i \alpha})^2 \Big],
\ee
where $\psi_{R/L;i \alpha} \sim e^{i(\phi_{i \alpha}\pm \theta_{i \alpha})}$,
and the dual fields satisfy 
$[ \phi_{i \alpha}(x), \theta_{j \beta}(y)]=-i \pi \delta_{ij}
 \delta_{\alpha \beta}\,\Theta(x-y)$. $\Theta(x)$ is a Heaviside step function.
 The externally screened Coulomb interaction is
 \be
 H_{{\rm int}}=e^2 \ln(R_s/R) \int dx \big(\sum_{i \alpha} \rho_{i \alpha}\big)^2,
 \ee
 where $ \ln(R_s/R)$ is a parameter of carbon nanotubes.
 Since the $H_{{\rm int}}$ invovles only the total charge density
 $\sum_{i \alpha} \rho_{i \alpha}$, it is desirable to introduce a 
 charge and spin channel  to manifest the underlying symmetry.
 \be
 \label{def1}
 \theta_{i,\rho/\sigma}=(\theta_{i \uparrow} \pm \theta_{i
 \downarrow})/\sqrt{2},
 \ee with similar definitions for $\phi$.  
 Define also
 \be
 \label{def2}
\theta_{\mu,\pm}=(\theta_{1 \mu}\pm \theta_{2 \mu})/\sqrt{2}, \mu=\rho/\sigma
\ee
again with similar definitions for $\phi$.
As defined in Eq.(\ref{def1}, \ref{def2}), the 
new fields $\theta_a,\phi_a$ with $a=(\rho/\sigma,\pm)$, satisfy the 
canonical commutators $[ \phi_a(x), \theta_b(y)]=-i \pi \delta_{ab}\,
\Theta(x-y)$. Now the total Hamiltonian $H=H_0+H_{{\rm int}}$ can be expressed
in terms of new fields $\theta_a, \phi_a$ as follows
\ba
H&=&\sum_a H_a,\nonumber \\
H_{a}&=&\int dx \frac{v_{a}}{2\pi} \Big[
g^{-1}_{a} (\partial_x \theta_{a})^2+ 
g_{a} (\partial_x \phi_{a})^2 \Big],
\ea
\ba
v_{\rho +}&=&\big[ v_F(v_F+ (8e^2/\pi \hbar) \ln(R_s/R))\big]^{1/2} >
v_{\sigma +}, \nonumber \\
g_{\rho +}&=&v_F/v_{\rho +}.
\ea
\be
\label{symmetry}
v_{\rho -}=v_{\sigma +}=v_{\sigma -}=v_F, 
g_{\rho -}=g_{\sigma +}=g_{\sigma -}=1.
\ee
The Eq.(\ref{symmetry}) specifies the symmetry  of FSCM. For generic 
interaction, such a symmetry is not expected.
We   have deliberately written the  Hamiltonian in a 
more general form to take other cases into account.
In the following, we will use the general notations assuming 
no special relations like  Eq.(\ref{symmetry}). When the relation 
Eq.(\ref{symmetry}) plays a crucial role, such a case will be treated separately.
The propagators of boson fields in imaginary frequency are
\ba
\label{propagators}
\langle \theta_a(k,i\omega)\,\theta_b(-k,-i\omega) \rangle&=&\delta_{ab}
\frac{\pi v_a g}{v_a^2 k^2+\omega^2}, \nonumber \\
\langle \phi_a(k,i \omega)\,\phi_b(-k,-i \omega) \rangle&=&\delta_{ab}
\frac{\pi v_a/ g}{v_a^2 k^2+\omega^2}, \nonumber \\	
\langle \theta_a(k,i\omega)\,\phi_b(-k,-i\omega) \rangle&=&
\langle \phi_a(k,i\omega)\,\theta_b(-k,-i\omega) \rangle  \nonumber \\
&=&	\delta_{ab}
\frac{-i \pi k \omega}{(v_a^2 k^2+ \omega^2) k^2}.
\ea
The propagators Eq.(\ref{propagators}) are all the information we need to
compute the Raman cross-section.
%%%%%%%%%%%%%%%%%%%%%%%%%%%%%%%%%%%%%%%%%%%%%%%%%%%%%%%
Next, we recall some basic facts of (resonant) Raman spectroscopy. 
The differential  cross-section of resonant Raman scattering 
is given by \cite{kramer,ramantheo}
\ba
\frac{d^2 \sigma}{d \Omega d \omega}&=&
\left( \frac{e^2}{m c^2} \right)^2\,
\frac{\omega_F}{\omega_I}\,\frac{n_\omega+1}{\pi}\,{\rm Im} \chi(\vq,\omega),
\nonumber \\
\chi(\vq,t)&=& i \theta(t)\,\langle [ N(-\vq,t),N(\vq,0) ] \rangle,\nonumber \\
N(\vq)&=&\sum_{i,\alpha,\vk}\,  \frac{\gamma_\alpha}{D(\vk)} 
c^\dag_{i \alpha}(\vk+\vq)\,c_{i \alpha}(\vk),
\ea
where
$n_\omega=\frac{1}{e^{\beta \omega}-1},
\vq=\vk_I-\vk_F,\;\;\omega=\omega_I-\omega_F$.
$c_{i \alpha}(\vk)$ is an i-th electron
operator with momentum $\vk$ and spin $\alpha$, and 
$\vk_{I(F)}, \omega_{I (F)},{\bf e}_{I (F)}$ 
are the momentum, energy, and the polarization vector of the incident (scattered)
photons, respectively. 
 The electron operator $c_{i \alpha}(\vk)$ 
can decomposed into  the chiral fermions $\psi_{R/L;i \alpha}$ of Eq.(\ref{h0}). 

The coefficents of the scattering operator $N(\vq)$ are given by
\ba
\label{coeff}
\gamma_\alpha&=&\gamma_0 {\bf e}_I \cdot {\bf e}_F
+ i \alpha \gamma_1 \,|{\bf e}_I \times {\bf e}_F
|,\nonumber \\
D(\vk)&=& E_c(\vk+\vq)-E_v(\vk+\vq-\vk_I)-\hbar \omega_I.
\ea
$E_c$ and $E_v$ are the conduction and valence band energies. 
$\gamma_0$ is  the transition matrix element
between valence and conduction band, which is assumed to be constant.
The spin-dependent term $\gamma_1$ of scattering operator $N(q)$ originates
from the spin-orbit coupling \cite{ramantheo}. The spin-orbit 
coupling is supposed to  small in carbon nanotube system, therefore,  
practically the spin-dependent scattering may be  difficult to observe 
in actual carbon nanotube system.
%%%%%%%%%%%%%%%%%%%%%%%%%%%%%%%%%%%%%%%%%%%%%%%%%%%%%%%%%%%%%%%%%%%%%%

Away from the resonance, namely, $\hbar v_F q << |E_G-\hbar \omega_I|$ and 
$\hbar \omega << |E_G-\hbar \omega_I|$, the momentum dependence of $D(\vk)$ can 
be neglected. Here, $E_G$ is the distance between the conduction and 
valence band at the Fermi point.
Then the scattering operator $N(\vq)$ can be simplified to \cite{kramer}
\ba
\label{classical}
N(\vq)&=&\frac{1}{E_G-\hbar \omega_I}\,
\Big[\gamma_0 {\bf e}_I \cdot {\bf e}_F \,\rho(\vq) +
     i\gamma_1|{\bf e}_I \times {\bf e}_F | \sigma(\vq) \Big],\nonumber \\
\rho(\vq)&=&\sum_{i=1,2} \big(\rho_{i \uparrow}(\vq)+ \rho_{i \downarrow}(\vq)
\big),\nonumber \\
\sigma(\vq)&=&\sum_{i=1,2}\big( \rho_{i \uparrow(\vq)}- \rho_{i \downarrow}(\vq)
\big).
\ea
Depending on the relative polarizations (parallel or perpendicular) of the incident
and scattered photons, only the charge or spin channel contributes 
to the Raman cross section, respectively.
This is  the {\em classical } selection rule of Raman spectra of quantum wires 
and dots, which is valid in the lowest order of 
$\frac{\hbar v_F q }{|E_G-\hbar \omega_I|},
\frac{\hbar \omega }{|E_G-\hbar \omega_I|}$.
Close to the
resonance such that $\hbar \omega_I \approx E_G+ \hbar v_F q$,
the momentum dependence of $D(k)$ should be taken into account. 
Following Sassetti and Kramer
\cite{kramer,kramer2,comment}
we expand the  $\frac{1}{D(k)}$ to include the momentum dependence (assuming
back scattering $k_i=q/2$ and decomposing the electron into the 
chiral fermions), and
we obtain two leading  corrections to the scattering operator $N(q)$.  
\ba
\label{op1} 
& &\delta N_1^{(r)}(q)=-\frac{  \eta_1}{(E_G-\hbar \omega_I)^2}\,\sum_{i,\alpha,k}\,
 \nonumber \\
& &\times \gamma_\alpha \,v_{F}\,(r k - k_{F})
 \psi_{i \alpha}^{(r)\dag }(k+q)\,\psi_{i \alpha}^{(r)}(k), \nonumber \\
& &\delta N_2^{(r)}(q)=-\frac{ \eta_2 v_F q}{(E_G-\hbar \omega_I)^2}\,
\sum_{i, \alpha,k}\,\nonumber \\
& &\times \gamma_\alpha 
 \psi_{i \alpha}^{(r)\dag }(k+q)\,\psi_{i \alpha}^{(r)}(k), 
\ea
where $r=R/L=\pm1$ and $\eta_1, \eta_2$ are dimensionless numbers determined by
the detailed shape of bands.
The operators $\delta N_{1,2}^{(r)}(q)$ can be  expressed in terms of
density operators $\rho_{R/L;i\alpha}$ by using the commutation relation
$\big[ \rho_{r s}(q),\rho_{r s}(q^\prime) \big]=
-\frac{r q L}{2\pi}\,\delta_{q+q^\prime,0}$, and then can be bosonized.
The bosonized form of $\delta N_{1,2}(q)=\sum_r \delta N_{1,2}^{(r)}(q)$
are given by 
\ba
& &\delta N_1(q)=\frac{- \eta_1 2\pi v_F}{L (E_G -\hbar \omega_I)^2}
\sum_p\, \big[ \frac{ ip}{2\pi} \frac{i(q-p)}{2\pi} \big]
\Big[ \gamma_0 {\bf e}_I \cdot {\bf e}_F \nonumber \\
& & \times:\Big(
\theta_{\rho +}(p) \theta_{\rho +}(q-p) 
+ \theta_{\sigma +}(p) \theta_{\sigma +}(q-p) \nonumber \\
& &+\theta_{\rho -}(p) \theta_{\rho -}(q-p)+ \theta_{\sigma -}(p) \theta_{\sigma -}(q-p)
  +(\theta \to \phi) \Big): \nonumber \\
 & &+ i \gamma_1 |{\bf e}_I \times {\bf e}_F|  
 :\Big(
\theta_{\sigma +}(p) \theta_{\rho +}(q-p) \nonumber \\
& &+ \theta_{\rho +}(p) \theta_{\sigma +}(q-p)+
\theta_{\sigma -}(p) \theta_{\rho -}(q-p) \nonumber \\
& &+ \theta_{\rho -}(p) \theta_{\sigma -}(q-p)
+(\theta \to \phi) \Big) : \Big].
\ea
where $: :$ denotes the normal ordering with respect to the filled Fermi sea.
\be
\delta N_2(q)=\frac{ \eta_2 v_F q}{ (E_G-\hbar \omega_I)^2} \frac{2 i q}{\pi}\,
\Big[ \gamma_0 \theta_{\rho +} + i \gamma_1 \theta_{\sigma +} \Big].
\ee
In passing we note that the correlation function 
$\langle \delta N_1(i\omega,q) \delta N_2(-i \omega,q) \rangle$ 
vanishes because it is the expectation
value of {\it odd} powers of boson fields $\theta, \phi$, while the Lagrangian 
is even in $\theta,\phi$.
The correlation function 
$\langle \delta N_2(q,i\omega)  \delta N_2(-q,-i\omega) \rangle$ does not 
yield interesting features, so we focus on the 
$\langle \delta N_1(q,i\omega)  \delta N_1(-q,-i \omega) \rangle$.
In actual computation we use the imaginary time formulation, and later 
analytically continue to the real frequency $
{\rm Im} \chi^R(q,\omega)={\rm Im} \chi(q, i\omega \to \omega+i 0^+)$.
At this point, we are ready to compute the correlation function
${\rm Im} \chi^R(q,\omega)$ with a correction near resonance included.
%%%%%%%%%%%%%%%%%%%%%%%%%%%%%%%%%%%%%%%%%%%%%%%%%%%%%%%%%%%%
\section{Results and Discussions}
The calculation of correlation function is rather tedious but straightforward.
We do not show the details of calculation, but focus on the results.
The correlation function are computed at zero temperature because 
most of our essential results do not depend on temperature.

First, consider the polarized spectra with
${\bf e}_I \cdot {\bf e}_F=1, {\bf e}_I \times {\bf e}_F=0$.
In the lowest order of the $\frac{\hbar v_F q}{E_G-\hbar \omega_I}$, 
only the total charge  density operator contributes to the correlation function 
$\chi(q,\omega)$ (See Eq.(\ref{classical})).  
The correlation function can be computed easily.
\be
\label{lowest}
{\rm Im} \chi_{{\rm pol}}^{(0)}(q,\omega) \sim 
\frac{L \,\gamma_0^2\,q}{(E_G-\hbar \omega_I)^2}\,
g_{\rho +}\delta(\omega- v_{\rho +} q),
\ee 
As expected, a peak corresponding to the total charge collective 
excitation appears with a dispersion relation
$\omega=v_{\rho +} q$.
Near resonance, the contribution from the correction $\delta N_1$ 
should be included. It is given by
\ba
\label{pol}
& &{\rm Im}  \chi_{{\rm pol}}^{(1)}(q,\omega) \sim
\frac{ L \eta_1^2 \gamma_0^2 v_F^2 }{
(E_G-\hbar \omega_I)^4}\Big[ \nonumber \\
& &\sum_a (g_a^2+g_a^{-2}+2)\, \frac{q^3}{6}\,\delta(\omega-v_a q) \nonumber \\
& &+\sum_a  \Theta(\omega-v_a q) \frac{(g_a^2+g_a^{-2}-2) }{2 v_a}
\big( (\omega/v_a)^2-q^2 \big)\Big].
\ea
The most notable feature of Eq.(\ref{pol}) is the appearance of
 the peaks corresponding to
{\it all} of the collective excitations (the first term in the bracket).
The last term in the bracket is some background negligible near the peak, 
which was 
not explicitly evaluated by Sassetti and Kramer.
The dependence the peaks on the photon energy and the momentum  are
 identical with
those  of Sassetti and Kramer's results on the single-channel case. 
Interestingly, the 
background {\em disappears for non-interacting electrons}, i.e. all $g_a=1$.
Thus, the magnitude of the background can give some measure of interactions.
At finite temperature, the peak intensity $q^3$ of (\ref{pol}) becomes
$q(q^2+T^2/v^2_{\sigma +})$, which is also identical with that of single-channel
case\cite{comment}.

Next consider the depolarized spectra with
${\bf e}_I \cdot {\bf e}_F=0, |{\bf e}_I \times {\bf e}_F|=1$.
In the lowest order of 
$\frac{\hbar v_F q}{E_G-\hbar \omega_i}$, only the total 
spin operator contributes to the correlation
function in accordance with the classical selection rule (\ref{classical}).
\be
{\rm Im} \chi^{(0)}_{{\rm dep}}(q,\omega) \sim
\frac{L\,\gamma_0^2\,q}{(E_G-\hbar \omega_I)^2}\,
g_{\sigma +} \delta(\omega-v_{\sigma +} q ).
\ee
Near resonance, the contributions from the correction $\delta N_1$ 
should be included. Recall that we did not impose any special relations
among velocities $v_a$ and Luttinger parameters $g_a$.
The contribution reads ($\omega, q >0$)
\ba
\label{depol}
& &{\rm Im}  \chi_{{\rm dep}}^{(1)}(q,\omega) \sim 
\frac{L v_F^2 \gamma_1^2 
\eta_2^2}{8 (E_G-\hbar \omega_I)^4} \Big[ \nonumber \\ 
& & G_1 \Theta(\omega-v_{\sigma +} q) \Theta( q v_{\rho +} -\omega)
(\omega-v_{\sigma +} q) ( q v_{\rho +} -\omega)
 \nonumber \\
& & +G_2 \Theta(\omega-v_{\sigma +} q)
  (\omega+ v_{\rho +} q) (\omega-v_{\sigma +} q) \nonumber \\
 & &+ G_2 \Theta(\omega-v_{\rho +} q) 
(\omega+ v_{\sigma +} q)(\omega-v_{\rho +} q) \nonumber \\ 
&+&(g_{\rho/\sigma +}, v_{\rho/\sigma +} \to
g_{\rho/\sigma -}, v_{\rho/\sigma -})\Big],
\ea
where $G_1=\frac{(g_{\rho +} g_{\sigma +} +g_{\rho +}^{-1} g_{\sigma +}^{-1}+2)}{
(v_{\rho +}-v_{\sigma +})^3},
G_2=\frac{(g_{\rho +} g_{\sigma +} +g_{\rho +}^{-1} g_{\sigma +}^{-1}-2)}{
(v_{\rho +}+v_{\sigma +})^3}$.
For generic values of $v_a, g_a$ the incoherent background exists
starting from $v_{\sigma +} q$ (we assume $v_{\rho +} > v_{\sigma +}$ as 
in carbon nanotube systems).
The second and the third term of Eq.(\ref{depol}) vanish for 
non-interacting electrons ($G_2 \to 0$) as in the polarized spectra.
Thus, for generic interactions the overall feature of the resonant depolarized
spectra of two-channel quantum wire is similar to 
that of single-channel quantum wire.

In the case of FSCM,  there are special relations 
among velocities and Luttinger 
parameters specified by Eq.(\ref{symmetry}). 
Since $v_{\rho -}=v_{\sigma -}$, the following term  Eq.(\ref{singular}) of 
Eq.(\ref{depol}) becomes 
singular and should be reconsidered. For other parts,
 we can simply substitute  $v_{\rho -}=v_{\sigma -}$.
\ba
\label{singular}
& &\frac{(g_{\rho -} g_{\sigma -} +g_{\rho -}^{-1} g_{\sigma -}^{-1}+2)}{
(v_{\rho -}-v_{\sigma -})^3}\, \nonumber \\
&  &\times \Theta(\omega-v_{\sigma -} q) \Theta( q v_{\rho -} -\omega)
(\omega-v_{\sigma -} q) ( q v_{\rho -} -\omega).
\ea 
When Eq.(\ref{singular}) is treated carefully in the limit\cite{comment2}
$v_{\rho -} \to v_{\sigma -}, g_{\rho -} \to g_{\sigma -}$, it becomes
\be
\frac{ q^3}{6} ( g_{\rho -}^2+g_{\rho -}^{-2}+2)\,
\delta(v_{\rho -}q-\omega).
\ee
The above result implies that for the two-channel quantum wires with
special symmetry among interactions, the "SPE" peak can also appear 
in the depolarized spectra
with the same momentum and photon energy dependences 
as in the polarized spectra.

In summary, we have studied the resonant Raman spectra of two-channel
quantum wires. In resonant polarized  spectra, {\em all} the collective 
excitations of the system (not just charge and spin) give rise to the 
"SPE" peaks. For generic interactions without special symmetry among
interactions, no "SPE" peak appear in the resonant {\it depolarized} spectra.
But if a certain symmetry makes  two velocities and Luttinger parameters
coincide then the "SPE" peak can emerge in the depolarzied spectra, too.
The armchair (N,N) carbon nanotube systems 
does have such a symmetry, and SPE peak in depolarized spectra is expected.
Experimental verifications of 
the existence of SPE peak in depolarized configuration in nanotubes
or in other two-channel quantum wires are proposed.

%%%%%%%%%%%%%%%%%%%%%%%%%%%%%%%%%%%%%%%%%%%%%%%%%%%%%%%%%%
\bigskip
\centerline{{\bf ACKNOWLEDGEMENTS}}
We are grateful to H. Y. Choi and H. W. Lee for useful discussions.
This work was  supported by the Korea Science and Engineering
Foundation (KOSEF) through the grant No. 1999-2-11400-005-5, and by the 
Ministry of Education through Brain Korea 21 SNU-SKKU Program.

%%%%%%%%%%%%%%%%%%%%%%%%%%%%%%%%%%%%%%%%%%%%%%%%%%%%%%%%%%%%%%%%%%%%%%

%%%%%%%%%%%%%%%%%%%%%%%%%%%%%%%%%%%%%%%%%%%%

\end{multicols}

\begin{references}
\bibitem{comment} We have neglected the effect of finite range of interactions.
For the discussion on the interactions with finite range, see the work by
Kramer and Sassetti\cite{kramer2}.
\bibitem{comment2} In case $v_{\rho -}=v_{\sigma -}$, it is simplest to calculate
the correlation function from the outset assuming such relation.
\end{references}
\end{document}